%% file: main.tex
\documentclass[%
 reprint,
nofootinbib,
 amsmath,amssymb,
 aps,
 prl,
]{revtex4-2}

\usepackage{graphicx}
\usepackage{dcolumn}
\usepackage{bm}

\input{preamble}

\begin{document}

\title{Quantifying non-invertible chiral symmetry breaking}

\author{Finn Gagliano}
 \email{finn.gagliano@durham.ac.uk}
\affiliation{%
 Department of Mathematical Sciences, Durham University, UK
}%

\begin{abstract}
It has recently been suggested that the chiral symmetry of the Standard Model is not explicitly broken by its ABJ anomaly, but is instead made non-invertible. In this work, I quantify the breaking of non-invertible symmetries of this type by calculating the symmetry breaking scales and effective charges $q(r)$ originally studied for continuous invertible higher-form symmetries. As an example I consider the explicit breaking of the chiral symmetry of massless QED by introducing Yukawa couplings between the electron and a massive real scalar in the UV. I briefly comment on how one might quantify the breaking of general non-invertible symmetries.
\end{abstract}

\maketitle


\section{Introduction} \label{sec:intro}

The past decade has seen a major change in the way a symmetry of a quantum field theory is defined. Building on the approach of defining an ordinary symmetry as a topological codimension-1 operator, \cite{Gaiotto:2014kfa} defined a $p$-form symmetry, a type of \textit{generalised symmetry}, to be a topological codimension-\textit{$(p+1)$}  operator, acting on $p$-dimensional extended operators such as Wilson and 't Hooft defects. These symmetries exist in a diverse range of quantum field theories, including the Standard Model \cite{Tong:2017oea}. This has prompted a new way of studying the Standard Model, offering new perspectives on its structure, anomalies, and possible extensions \cite{Wang:2018jkc,Garcia-Etxebarria:2018ajm,Wan:2019gqr,Wang:2020xyo,Wang:2020gqr,Wang:2021hob,Anber:2021upc,Wang:2021vki,Wang:2021ayd,Wang:2022eag,Choi:2022rfe,Cordova:2022fhg,Choi:2023pdp,Agrawal:2023sbp,Cordova:2023her,Cordova:2024ypu,Li:2024nuo,Alonso:2024pmq,Koren:2024xof,Hsin:2024lya,Delgado:2024pcv,Wan:2024kaf,Gagliano:2025gwr,Chen:2025buv,Hamada:2025cwu,Alonso:2025rkk}. See \cite{Brennan:2023mmt,Iqbal:2024pee,Gomes:2023ahz,Luo:2023ive,Shao:2023gho,Loaiza-Brito:2024hvx,Cordova:2022ruw,Schafer-Nameki:2023jdn,Bhardwaj:2023kri,Costa:2024wks,Reece:2023czb,Davighi:2025iyk} for a range of introductions to generalised symmetry.

The study of generalised symmetry also includes \textit{non-invertible} symmetries, which are no longer described by groups but by more general structures known as categories. Such symmetries have long been known to exist in 2d, but it was only in \cite{Kaidi:2021xfk,Choi:2021kmx} that examples were first found in 4d QFTs. Since then, non-invertible symmetries have been found in a variety of theories, again including the Standard Model \cite{Cordova:2022ieu,Choi:2022jqy}.

In this work, I will discuss the explicit breaking of the chiral symmetry of massless QED and how its breaking scale can be quantified. This symmetry has a pure 't Hooft anomaly obstructing its gauging, as well as an ABJ anomaly. In the past, symmetries with an ABJ anomaly were not considered to be valid symmetries of the quantum theory due to their current conservation being violated by quantum effects. As I will review shortly, \cite{Cordova:2022ieu,Choi:2022jqy} suggested that the ABJ anomaly does not break the chiral symmetry, but instead makes it \textit{non-invertible}\footnote{In \cite{Putrov:2023jqi} a similar stance was taken for symmetries with gravitational anomalies. While the chiral symmetry suffers from such an anomaly, the breaking mechanism we use in this work will also break non-invertible symmetries of this type, and so we can safely ignore any contribution from this anomaly.}.

For (non-invertible) higher-form symmetries, the standard way for the symmetry to be broken is to consider adding additional matter to the theory that screens the charged extended operators to the identity. For non-invertible 0-form symmetries, we must therefore add operators to the action that explicitly breaks the symmetry. Given that the Standard Model has such a symmetry, enumerating breaking mechanisms and their phenomenological upshots is useful for better understanding our universe. The approach I will take for massless QED is to add a Yukawa coupling between the electron and a massive real scalar in the UV that breaks the chiral symmetry at the level of the Lagrangian. The main aim of this work is to show how one can extend the methods of quantifying the explicit breaking of continuous invertible higher-form symmetries proposed in \cite{Cordova:2022rer} to non-invertible symmetries that arise in a similar way to the chiral symmetry.

\section{Non-invertible symmetry breaking scales}

A prominent class of non-invertible symmetries are given by so-called ABJ-like symmetries, where a would-be conserved current for a continuous $p$-form symmetry is violated:
\begin{equation} \label{eq:general-current-violation}
    d*j_{p+1} = V_{k+1} \wedge W_{d-p-k-1}
\end{equation}
where $V_{k+1},W_{d-p-k-1}$ are closed forms that obstruct us from obtaining an invertible symmetry. The operator we would define for the $p$-form symmetry,
\begin{equation}
    \cU_\alpha(\Sigma_{d-p-1}) = e^{i\alpha\int_\Sigma *j},
\end{equation}
is then non-topological - if we consider a smooth deformation of $\Sigma$ to some other $\Tilde{\Sigma}$ then we get
\begin{equation}
    \cU_\alpha(\Tilde{\Sigma}) - \cU_\alpha(\Sigma) = e^{i\alpha\int_{(\Delta\Sigma)_{d-p}}d*j_{p+1}} \neq 1
\end{equation}
so that correlation functions depend dynamically on $\Sigma$. Thus, this operator does not define a symmetry of our system.

However, suppose we can locally write
\begin{equation}
    V_{k+1} = dv_{k},
\end{equation}
or similar for $W$, so that we can rewrite \eqref{eq:general-current-violation} as
\begin{equation}
    d(*j_{p+1}-v\wedge W) = 0
\end{equation}
to obtain a modified conserved current. Then, we can define a \textit{topological} operator
\begin{equation} \label{eq:tilde-operator}
    \Tilde{\cU}_\alpha(\Sigma_{d-p-1}) = e^{i\alpha\int_\Sigma *j -v\wedge W}.
\end{equation}
On general $\Sigma_{d-p-1}$, $v_k$ might be subject to large gauge transformations
\begin{equation} \label{eq:large-gauge}
    v_k \rightarrow v_k + \lambda_k,\ \int \lambda_k \neq 0,
\end{equation}
so that \eqref{eq:tilde-operator} is not gauge invariant if $\int W \neq 0$. A way to circumnavigate this is to `stack' a $(d-p-1)$-dimensional TQFT $\cA_\alpha(\Sigma)$ onto the operator, with $\cA_\alpha$ transforming in the opposite way under gauge transformations \cite{Cordova:2022ieu,Choi:2022jqy}:
\begin{equation}
    \widehat{\cU}_{\alpha}(\Sigma) = \cA_\alpha \otimes \Tilde{\cU}_\alpha(\Sigma)
\end{equation}
so that this is then a topological gauge-invariant operator, and therefore defines a valid symmetry of our theory. However, the minimal choice of $\cA_{\alpha}$ can in general be a \textit{non-invertible} TQFT, i.e. there is no other TQFT that we can multiply this by to obtain the identity. Thus, symmetries of this type become non-invertible due to the stacking with non-invertible TQFTs.

\subsection{Quantifying non-invertible symmetry breaking}

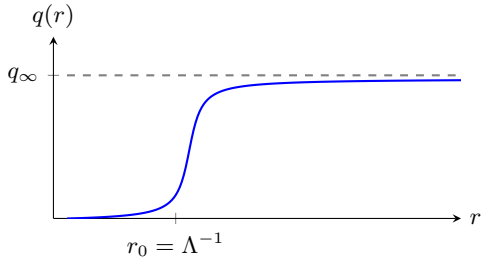
\begin{figure}[t]
\begin{center}
    \begin{tikzpicture}
    \begin{axis}[
        axis lines = middle,
        xlabel = \(r\),
        ylabel = {\(q(r)\)},
        domain=0.1:3,
        samples=200,
        ymin=0, ymax=0.8,
        xmin=0, xmax=3,
        xtick={0.9},
        ytick={0.63},
        xticklabels={\( r_0 = \Lambda^{-1} \)},
        yticklabels={\(q_\infty\)},
        width=7cm,
        height=4cm,
        every axis y label/.style={at={(current axis.above origin)}, anchor=south},
        every axis x label/.style={at={(current axis.right of origin)}, anchor=west},
      ]
    \addplot [blue, thick] {0.011/pi *atan(15*(x - 1)) + 0.3};
    \addplot [dashed, thick, gray] {0.63};
    \end{axis}
    \end{tikzpicture} 
\end{center}
\caption{Effective charge $q(r)$ varying with the radius. Symmetry is badly broken at $r_0=\Lambda^{-1}$ as $q(r_0) \approx 0$, and at large $r$ tends to the IR charge $q_\infty$.}
\label{fig:graph}
\end{figure}

Given a continuous invertible $p$-form symmetry, we will have an ordinary topological operator
\begin{equation}
    \cU_\alpha(\Sigma_{d-p-1}) = e^{i\alpha \int *j_{p+1}}
\end{equation}
that generates the symmetry. Suppose that the symmetry is explicitly broken at some UV scale $\Lambda$. A way of quantifying this breaking scale was described in \cite{Cordova:2022rer}, and goes as follows. In the IR, the symmetry will act on a charged operator $\cW_q(\gamma_p)$ by
\begin{equation} \label{eq:sym-action}
    \braket{\cU_\alpha(\Sigma) \cW_q(\gamma) \dots} = e^{i\alpha q} \braket{\cW_q(\gamma)\dots}
\end{equation}
where we assume that $\Sigma$ and $\gamma$ have unit linking number. We can let $\Sigma = S^{d-p-1}_r$, where $r$ is the radius of the sphere we use to measure the charge $q$. Then, as $r\approx1/\Lambda$, we expect the symmetry to be explicitly broken such that $\cU_\alpha$ acts trivially on the charged operator, meaning $q=0$ at these length scales. Following \cite{Cordova:2022rer}, this can be seen as letting the charge depend on $r$:
\begin{equation} \label{eq:effective-charge}
    q(r) \coloneqq \lim_{\alpha\rightarrow 0}\frac{1}{i\alpha} \cdot \log \frac{\braket{\bar{\cW}_q(\gamma^\infty) \cU_\alpha(S^{d-p-1}_r) \cW_q(\gamma^0)}}{\braket{\bar{\cW}_q(\gamma^\infty) \cW_q(\gamma^0)}}
\end{equation}
where $\gamma^\infty$ is some $p$-dim curve at infinite distance from $\gamma^0$, the same curve but placed at the origin. Similarly, we can describe this effective charge as
\begin{equation} \label{eq:delta-effective-charge}
    q(r) = q_\infty (1 - \delta q(r)) 
\end{equation}
where $\delta q(r)$ is the correction to the numerator of \eqref{eq:effective-charge} from symmetry-breaking effects and $q_\infty$ is the constant IR charge of \eqref{eq:sym-action}.
We then say that the symmetry is broken below some length scale $r_0=1/\Lambda$ if
\begin{equation} \label{eq:breaking-condition}
    \frac{1}{q_\infty}\left.\frac{d(\delta q(r))}{d\log r}\right\vert_{r=r_0} \approx \cO(1).
\end{equation}
We can picture this condition for symmetry breaking via FIG. \ref{fig:graph}. That is, we require
\begin{equation} \label{eq:vanishing-charge}
    q(r_0) \approx 0
\end{equation}
for the symmetry to be explicitly broken above energy scales $r_0^{-1}=\Lambda$.

Note that the discussions of effective charge above hold for a \textit{continuous invertible} symmetry: taking the appropriate limit in \eqref{eq:effective-charge} requires continuous $\alpha$, and we also require the symmetry to act on charged operators as in \eqref{eq:sym-action}, i.e. with an invertible phase. However, it has been pointed out in recent works \cite{Karasik:2022kkq,GarciaEtxebarria:2022jky} that certain values of $\alpha$ result in $\cA_\alpha = 0$ when $\int W \neq 0$, such that the non-invertible symmetry $\widehat{\cU}_\alpha$ acts as a projector, similarly to the more recent work \cite{Jacobson:2024muj}, rather than with a phase. In the case of massless QED, \cite{Cordova:2022ieu,Choi:2022jqy} constructed non-invertible symmetries for $\alpha \in \bQ / \bZ$ by stacking with a TQFT, in analogy with the fractional quantum Hall effect. However, \cite{Karasik:2022kkq} showed that $\cA_\alpha$ vanishes unless
\begin{equation}
    \alpha \int dA \in \bZ . 
\end{equation}
Additionally, \cite{Karasik:2022kkq,GarciaEtxebarria:2022jky} proposed an alternative TQFT, describing a compact boson defined only on $\Sigma$, that preserves the full $U(1)_\chi$ symmetry. Nonetheless, the same vanishing condition still applies in this case.

So why do we expect the above quantification methods to still be applicable for the ABJ-like symmetries if these operators sometimes vanish and/or become discrete? Well, when considering $\Sigma = S^{d-p-1}_r$ we have that $\cU_\alpha(S^{d-p-1}_r)$ \textit{is} gauge-invariant and non-vanishing for all $\alpha$, as we have no large gauge transformations on $S^{d-p-1}_r$. This is because 
\begin{equation}
    H^k(S^{d-p-1}_r;\bZ) = 0\ \forall k=1,...,d-p-2
\end{equation}
so that there are no integral $\lambda \in H^k(S^{d-p-1}_r;\bZ)$ to shift $v_k$ by as in \eqref{eq:large-gauge}. We then don't have to stack the vanishing $\cA_\alpha$, so that we can compute the charge as normal using \eqref{eq:tilde-operator}. This also allows us to safely consider infinitesimal $\alpha$ on such backgrounds. Additionally, the ABJ-like symmetries we are considering act on charged operators as in \eqref{eq:sym-action} on such backgrounds, that is, with an invertible phase. Therefore, we can see that the definition of the effective charge in \eqref{eq:effective-charge} still holds for an operator of this form!

\section{QED chiral symmetry}

I now wish to consider the example of the $U(1)_\chi$ symmetry in massless QED. Consider the 4d Lagrangian
\begin{equation} \label{eq:lagrangian}
    \cL = \frac{1}{2e^2} F\wedge * F + i \bar{\psi}\slashed{D}\psi 
\end{equation}
where $\psi$ is the unit charge electron, represented as a Dirac fermion. The Lagrangian is invariant under an invertible chiral $U(1)_\chi$ transformation:
\begin{equation}
    \psi \xrightarrow{U(1)_\chi} e^{i\alpha \gamma_5} \psi,\ \bar{\psi} \xrightarrow{U(1)_\chi} \bar{\psi}e^{-i\alpha \gamma_5},
\end{equation}
which acts on left and right-handed fermions with opposite sign. The current for this symmetry is
\begin{equation}
    j_\mu = \bar{\psi}\gamma_\mu\gamma_5\psi
\end{equation}
and it is conserved classically. However, when we quantise the theory, the path integral measure is \textit{not} invariant under the symmetry due to instanton effects, and so the current conservation is violated in the quantum theory:
\begin{equation} \label{eq:current-violation}
    d*j_1 = F\wedge F.
\end{equation}
The would-be symmetry operator for the original current is given as
\begin{equation}
    \cU_\alpha(\Sigma_3) = e^{2\pi i\alpha\int_\Sigma *j}
\end{equation}
which is not topological due to the current violation. However, locally we can write the above current violation as
\begin{equation} \label{eq:trick}
    d(*j - A\wedge dA) = 0
\end{equation}
such that we can instead define a topological operator
\begin{equation} \label{eq:non-inv-operator}
    \Tilde{\cU}_\alpha = e^{2\pi i\alpha \int *j -A\wedge dA}.
\end{equation}
However, this is not invariant under large gauge transformations: $A\rightarrow A + \lambda$ where $\int \lambda =N \in \bZ$ gives a variation
\begin{equation}
    \cU_\alpha \rightarrow e^{2\pi i\alpha N} \cU_\alpha
\end{equation}
so that this is not gauge invariant unless $\alpha = \frac{p}{N}$ for $p \in \bZ$. This is precisely the issue regarding the $\bQ/ \bZ$ symmetry that we discussed qualitatively in the previous section. To retain the whole $U(1)_\chi$, we can stack \cite{Karasik:2022kkq,GarciaEtxebarria:2022jky}
\begin{equation}
    \cA_\alpha = \int [\cD \theta]e^{2\pi i\alpha \int d\theta \wedge dA}
\end{equation}
where $\theta$ is a compact boson with $d\theta$ transforming in the same way as $A$ under large gauge transformations, thus cancelling the gauge variance of $\Tilde{\cU}_\alpha$. As discussed above, it was shown in \cite{Karasik:2022kkq,GarciaEtxebarria:2022jky} that this vanishes unless $\alpha \int dA = 0 $. If we have $H^1(\Sigma_3;\bZ) = 0$ then there is no need to stack this TQFT, as discussed in the previous section. Further analysis of this continuous non-invertible symmetry can be found in \cite{Arbalestrier:2024oqg}.

\subsection{Coupling to massive particles}

I now wish to discuss a specific way to explicitly break the non-invertible chiral symmetry. A first point to make is that there is a very easy way to break the chiral symmetry - add magnetic monopoles \cite{Cordova:2022rer}. Doing so also means the magnetic 1-form symmetry is broken. This gives
\begin{equation}
    dF\neq 0
\end{equation}
which means we cannot go from \eqref{eq:current-violation} to \eqref{eq:trick}. Thus, we cannot define a topological operator for our $U(1)_\chi$ symmetry, so breaking the magnetic 1-form symmetry \textit{automatically} breaks the chiral symmetry. This sort of method gives us the inequality
\begin{equation}
    E_\chi \lesssim E_m
\end{equation}
where these are the energy scales above which the chiral and magnetic symmetries are broken, respectively. One could rephrase this as the energy scales at which these symmetries emerge under an RG flow from some UV completion of the theory, assuming these symmetries are spontaneously broken in the IR such that this notion of emergent symmetry is valid \cite{Cherman:2023xok}.

Adding monopoles is a breaking mechanism that gives us $E_\chi \approx E_m$. But what if we want 
\begin{equation}
    E_\chi \lnsim E_m \ ?
\end{equation}
It was pointed out in \cite{Cordova:2022fhg} that one can simply add an operator to the theory that classically breaks the symmetry at the level of the Lagrangian. I will study one such way to break the chiral symmetry at general $E_\chi$ by adding a Yukawa coupling between the electron and a massive real scalar in the UV that is neutral under the chiral symmetry:
\begin{equation} \label{eq:yukawa}
    \cL' \supset g\bar{\psi}\psi \phi + m^2\phi^2 + g'\phi^3
\end{equation}
where $\phi$ has mass $m$ so that at energy scales $\Lambda \ll E_\chi(m)$ we can integrate it out to obtain the Lagrangian of massless QED. The $\phi^3$ term guarantees that $\phi$ cannot be charged under the $U(1)_\chi$ symmetry, nor any finite subgroup\footnote{Examples given in \cite{Cordova:2022rer} broke the $U(1)$ symmetry to a finite subsymetry, which does not have a current and therefore did not impact their discussions. However, some finite charge would still remain in their example. We choose a real scalar to avoid this.}. Additionally, gauge invariance of the Yukawa term means that $\phi$ cannot be charged under the $U(1)$ gauge group. Finally, the Yukawa coupling explicitly breaks the symmetry at the level of the Lagrangian so that the current conservation is violated by the operator
\begin{equation}
    \partial_\mu j^\mu \approx g\bar{\psi}\psi\phi .
\end{equation}
Clearly, we cannot do the same trick as in \eqref{eq:trick} to obtain a non-invertible topological operator, as the right-hand side cannot be written as a total derivative. Note that this is the `minimal' coupling we can add - the mass dimension $[g]=0$ is marginal, and any higher-dimension coupling to the electron would therefore be irrelevant.

For this mechanism to be relevant to chiral symmetry breaking, we are assuming
\begin{equation}
    m \lesssim E_m 
\end{equation}
so that magnetic monopoles have not already broken the chiral symmetry. Note that in the $U(1)_Y$ sector of the Standard Model, such a coupling is \textit{not} gauge invariant - the gauge symmetry acts on left- and right-handed fermions differently, meaning a mass term for the electron is not gauge invariant. To remedy this, one might try to make $\phi$ complex and carry some appropriate charge under the gauge symmetry, but then any term we add to force the $U(1)_\chi$ charge of $\phi$ to vanish will not be gauge invariant. Therefore, such a (relevant/marginal) coupling cannot be added to break the $U(1)_\chi$ symmetry. It thus seems that $E_\chi \approx E_m$ in (some UV completion of) the Standard Model.

I wish to make a brief comment regarding the swampland. The swampland conjectures \cite{Vafa:2005ui,Ooguri:2006in} \footnote{See \cite{Banks:2010zn,Rudelius:2024vmc} for symmetry-centric introductions to the swampland, as well as \cite{vanBeest:2021lhn,Rudelius:2024mhq,Harlow:2022ich,Palti:2019pca,Brennan:2017rbf,Grana:2021zvf,Reece:2023czb} for more general introductions and reviews.}, which posit consistency conditions for theories of quantum gravity, often relate the completeness of the spectrum hypothesis \cite{Polchinski:2003bq} to the necessary absence of global symmetries \cite{Banks:2010zn}. Recent work \cite{Rudelius:2020orz,Heidenreich:2021xpr,Choi:2022fgx} has extended this to include generalised and non-invertible symmetries as well. For further examples involving non-invertible symmetry breaking and completeness, see \cite{McNamara:2021cuo,Arias-Tamargo:2022nlf}. In the familiar case of deriving QED from a $U(1)$ gauge theory, the introduction of a unit-charge electron both completes the electric spectrum and breaks the electric 1-form symmetry. Interestingly, this process also introduces the non-invertible chiral symmetry. As previously discussed, one can similarly complete the magnetic spectrum by adding dynamical monopoles, which breaks the magnetic 1-form symmetry and all other remaining global generalised symmetries, including the non-invertible chiral symmetry. This emphasises that even after completing the electric spectrum, residual `electric' global symmetries can persist, such as the non-invertible chiral symmetry, due to non-trivial mixing with magnetic 1-form symmetries. I expect a similar interplay may exist for flavour symmetries, which also mix with magnetic 1-form symmetries \cite{Cordova:2022qtz}. Similar examples have been discussed in the context of discrete gauge theories, twist vortices, and axion physics \cite{Rudelius:2020orz,Heidenreich:2021xpr,Choi:2022fgx}.

\subsection{Quantifying the breaking scale}

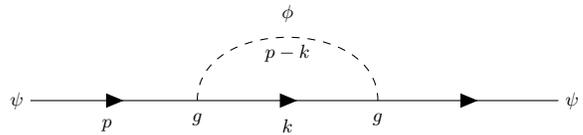
\begin{figure}[t]
    \centering
    \begin{tikzpicture}[scale=0.8, transform shape]
  \begin{feynman}
    \vertex (a) at (0,0) {\(\psi\)};
    \vertex[right=3cm of a] (b);
    \vertex[right=3cm of b] (c);
    \vertex[right=3cm of c] (d) {\(\psi\)};
    \vertex[right=1.5cm of b] (e);
    \vertex[below=0.2 of e] {\( k \)};
    \vertex[above=1.2 of e] {\( \phi \)};
    \vertex[right=1.5cm of a] (f);
    \vertex[below=0.2 of f] {\(p\)};
    \vertex[below=0.1 of b] {\(g\)};
    \vertex[below=0.1 of c] {\(g\)};
        
    \diagram* {
      (a) -- [fermion] (b) -- [fermion] (c) -- [fermion] (d),
      (b) -- [scalar, half left, looseness=1.2, edge label'=\(p - k\)] (c),
    };
  \end{feynman}
\end{tikzpicture}
    \caption{Correction to the electron propagator from Yukawa coupling $g\bar{\psi}\psi \phi$ that explicitly breaks the non-invertible chiral symmetry.}
    \label{fig:sunrise}
\end{figure}

I now wish to compute $q(r)$ for $U(1)_\chi$. To compute the numerator of \eqref{eq:effective-charge}, we need to consider
\begin{equation}
    \int_{S^3_r}\braket{\bar{\psi} (*j - A\wedge dA) \psi}.
\end{equation}
Following \cite{Cordova:2022rer}, to compute $\delta q(r)$ we need to consider how the addition of the Yukawa coupling \eqref{eq:yukawa} contributes to the above correlation function, which arises naturally as the contribution to the fermion self-energy. This is given by the Feynman diagram in FIG. \ref{fig:sunrise} and, leaving the details of all following loop calculations to the appendix, we get
\begin{equation} \label{eq:self-energy-initial}
    \Sigma(p) = g^2\int\frac{d^4k}{(2\pi)^4} \cdot \frac{1}{\slashed{k}} \cdot \frac{1}{(p-k)^2-m^2}
\end{equation}
which, assuming $p^2\ll m^2$, scales as
\begin{equation} \label{eq:self-energy-final}
    \Sigma(p) \approx \frac{g^2}{m^2} \cdot p^2 \cdot \slashed{p}.
\end{equation}

\noindent Then, we can obtain the 1-loop correction to the fermion propagator in momentum space as
\begin{equation}
    G(p) \approx \frac{i}{\slashed{p} - \Sigma(p)} = \frac{i\slashed{p}}{p^2(1-\frac{g^2}{m^2}p^2)}.
\end{equation}
To compute the effective charge, however, we require the position space propagator, so we make a Fourier transform:
\begin{equation}
    G(r) \approx \int d^4p \cdot e^{ip\cdot r} \cdot \frac{\slashed{p}}{p^2(1-\frac{g^2}{m^2}p^2)} .
\end{equation}
Then, this gives
\begin{equation}
    G(r) \approx -\delta q(r) G_0(r)
\end{equation}
for
\begin{equation}
    \delta q(r) \approx \frac{m^2r^2}{g^2} K_2\left(\frac{mr}{g} \right)
\end{equation}
where $K_2(s)$ is a Modified Bessel function of the second kind. As $s\ll 1 $, $K_2(s) \approx \frac{1}{s^2}$, so that when $r\ll \frac{g}{m}$ we get $\delta q(r) \approx 1$. For large $s$, $K_2(s) \approx \frac{1}{\sqrt{s}}e^{-s}$, so that when $r$ is very large $\delta q(r) \approx 0$ as it decays exponentially with $r$.

Therefore, by considering the form of $q(r)$ as in \eqref{eq:delta-effective-charge}, the energy scale $E_\chi = r_0^{-1}$ for which $q(r)$ obeys \eqref{eq:vanishing-charge} is clearly
\begin{equation} \label{eq:breaking-scale}
    E_\chi \approx \frac{m}{g} \ .
\end{equation}

So, we have been able to quantify the breaking of the non-invertible chiral symmetry as one would for a continuous invertible symmetry simply due to the behaviour of its topological operators when placed on a sphere.

\section{Future work}

In this work, I have focused on non-invertible $U(1)$ symmetries like the chiral symmetry of massless QED, and quantified their breaking scales by extending existing methods for continuous invertible higher-form symmetries. The symmetries we considered in this work have charges labelled by integers, but the possible charges under more general non-invertible symmetries are described by some appropriate representation category, see \cite{Lin:2022dhv,Bhardwaj:2023wzd,Bartsch:2023pzl,Bhardwaj:2023ayw,Bartsch:2023wvv,Copetti:2024onh,Cordova:2024iti,Copetti:2024dcz,GarciaEtxebarria:2024jfv,Choi:2024tri,Das:2024qdx,Bhardwaj:2024igy,Heymann:2024vvf,Choi:2024wfm}. These representation categories often account for how the non-invertible topological operators exchange charged particles from one `twisted sector' to another, or how they act on deep IR degrees of freedom, both of which we have not considered in this work. A categorical description of the effective charges under \textit{general} non-invertible symmetries could then be described by some smooth family of categories $\rep(r)$ that depends on the length scale $r$, compared to the fixed IR representations $\rep_\mathrm{IR}$ at large distance $r_{\infty}$:
\begin{equation}
    \rep(r_\infty) = \rep_\mathrm{IR}
\end{equation}
Then, we could define a property similar to \eqref{eq:vanishing-charge}:
\begin{equation}
    \rep(r_0)  = (d-1)\vect,
\end{equation}
where the UV representation category being trivial indicates that the symmetry is badly broken at length scales $r_0$. Such a description is most naturally defined in terms of a stack of tensor categories, and it would be interesting to explore this more general setup in future work.

\section*{Acknowledgments}
I would like to thank Bobby Acharya, Felix Christensen, Iñaki García Etxebarria, Jacob McNamara, Stefano Moretti, Tom Rudelius, and Christopher Tudball for helpful discussions. I would like to thank Iñaki García Etxebarria and Tom Rudelius for valuable comments on an earlier draft. I am funded by the STFC grant ST/Y509334/1.

\bibliography{refs}

\onecolumngrid
\appendix
\section{Appendix: Loop calculations for computing the effective charge in massless QED}

The aim of this appendix is to detail the computations used to quantify the breaking scale $E_\chi$ in massless QED. Firstly, however, I outline the logic of the computation to make it clear what the goal is. Rearranging \eqref{eq:effective-charge} after taking the limit, we have
\begin{equation}
    \int_{S^3_r} \braket{\bar{\psi}(*j -A\wedge dA)\psi } = q(r) \braket{\bar{\psi}\psi}.
\end{equation}
We want to compute the correction to the numerator of \eqref{eq:effective-charge} from symmetry-breaking effects, i.e. $\delta q(r)$, and so one way to do this is to write the RHS as 
\begin{equation}
    q(r) \braket{\bar{\psi}\psi} \approx q_\infty (G_0(r) +G_1(r) +...)
\end{equation}
where $G_1(r)$ are the 1st order loop corrections to $G_0(r) \approx \frac{\slashed{r}}{r^4}$, and then write
\begin{equation}
    G_1(r) \approx -\delta q(r) G_0 (r)
\end{equation}
so that the above becomes
\begin{equation}
    q(r) \braket{\bar{\psi}\psi} \approx q_\infty(1 - \delta q(r)) G_0(r)
\end{equation}
as expected. That is, to compute $\delta q(r)$, we need to compute the correction to the propagator, $G_1(r)$, and divide by $G_0(r)$. This is often a simpler computation than having to compute the LHS involving the current and, in this case, the photon.

We begin with the self-energy correction $\Sigma(p)$ in \eqref{eq:self-energy-initial}, and show how we can arrive at \eqref{eq:self-energy-final}. The first step is to write $\frac{1}{\slashed{k}} \approx \frac{\slashed{k}}{k^2}$ and then use the identity
\begin{equation}
    \frac{1}{AB} = \int_0^1dx \frac{1}{\left[ xA +(1-x)B\right]^2}
\end{equation}
where we take $A=(p-k)^2-m^2$ and $B=k^2$. This gives the integral
\begin{equation}
    \int_0^1dx \int d^4k \frac{\slashed{k}}{\left[k^2 -2x p\cdot k + xp^2 -xm^2 \right]^2}
\end{equation}
which, after changing coordinates to $\Tilde{k}=k-xp$, we get
\begin{equation}
    \int_0^1dx \int d^4\Tilde{k} \frac{\slashed{\Tilde{k}} +x\slashed{p}}{\left[\Tilde{k}^2 -\Delta \right]^2}
\end{equation}
where $\Delta = x(1-x)p^2 + xm^2$. Then, splitting this into two fractions, the one with numerator $\slashed{\Tilde{k}}$ vanishes by asymmetry by sending $\Tilde{k}\rightarrow-\Tilde{k}$. Then, we have
\begin{equation}
    \Sigma(p) \approx g^2 \slashed{p} \int_0^1dx \int d^4k \frac{x}{\left[ k^2 - \Delta \right]^2}.
\end{equation}
We can write the integral over $k$ as a standard dimensional regularisation result:
\begin{equation}
    \int d^4k \frac{1}{{\left[ k^2 - \Delta \right]^2}} \approx \frac{1}{\epsilon} + \log \Delta +\dots,
\end{equation}
which, ignoring the diverging term and assuming $p^2 \ll m^2$, gives
\begin{equation}
    \log \Delta \approx \log \left[ 1+ (1-x)\frac{p^2}{m^2}\right] + \log(xm^2).
\end{equation}
Ignoring the second term, which gives something scaling as $g^2 \log m^2 \cdot \slashed{p} $ after the $x$ integral and can therefore be ignored compared to \eqref{eq:self-energy-final}, we make a Taylor expansion
\begin{equation}
    \log \Delta \approx (1-x)\frac{p^2}{m^2}
\end{equation}
so that performing the integral over $x$ in $\Sigma(p)$ gives
\begin{equation}
    \boxed{ \Sigma(p) \approx \frac{g^2}{m^2} \cdot p^2 \cdot \slashed{p} }
\end{equation}
as in \eqref{eq:self-energy-final}.

Then, we can write the correction to the propagator, $G(p) = G_0(p) + G_1(p)$, as
\begin{equation}
    G(p) = \frac{i}{\slashed{p}-\Sigma(\slashed{p})} = \frac{i \slashed{p}}{p^2(1-\frac{g^2}{m^2}p^2)}.
\end{equation}
We can write this as the sum of two partial fractions
\begin{equation}
    G(p) = \frac{i\slashed{p}}{p^2} + \frac{i\frac{g^2}{m^2} \cdot \slashed{p}}{1-\frac{g^2}{m^2}p^2}
\end{equation}
so that we can clearly read off $G_1(p)$ as this second term.

The last step is to Fourier transform $G_1(p)$ into position space, as required:
\begin{equation}
    G_1(r) \approx \int d^4p e^{ir\cdot p} G_1(p) .
\end{equation}
If we let $M^2 = \frac{m^2}{g^2}$, then we can write this as
\begin{equation} \label{eq:g_1}
    G_1(r) \approx \frac{1}{M^2} \int d^4p e^{ir\cdot p} \frac{i\slashed{p}}{1-\frac{1}{M^2} p^2} = \int d^4p e^{ir\cdot p} \frac{i\slashed{p}}{M^2 - p^2} .
\end{equation}
Then, consider the form of the propagator for a scalar of mass $M$:
\begin{equation}
    G^\mathrm{scalar} (p) = \frac{i}{p^2 - M^2}.
\end{equation}
In position space,
\begin{equation}
    G^\mathrm{scalar} (r) \approx \int d^4p e^{ir\cdot p} \frac{i}{p^2 - M^2} \approx \frac{M}{r}K_1(Mr)
\end{equation}
where $K_1(s)$ is a modified Bessel function of the second kind. What we can see is that we can rewrite $G_1(r)$ as
\begin{equation}
    G_1(r) \approx \slashed{\partial} G^\mathrm{scalar} (r)
\end{equation}
to produce the $\slashed{p}$ in the numerator of \eqref{eq:g_1}, but then we can use the known form of $G^\mathrm{scalar} (r)$ in terms of the Bessel function to find $G_1(r)$. We then have that
\begin{equation}
    G_1(r) \approx \frac{M \slashed{r}}{r^3}  K_1(Mr) + \frac{M^2 \slashed{r}}{r^2}K_2(Mr)
\end{equation}
where we have used the property that
\begin{equation}
    \frac{d}{ds} K_n(s) = \frac{n}{s} K_n(s) + K_{n+1}(s) .
\end{equation}
Rewriting $G_1(r) \approx \delta q(r) G_0(r)$ where $G_0(r) \approx \frac{\slashed{r}}{r^4}$ then gives
\begin{equation}
    \delta q(r) \approx MrK_1(Mr) + M^2r^2K_2(Mr)
\end{equation}
and as both terms scale similarly for small and large $r$ we write for simplicity in the main text
\begin{equation}
    \boxed{ \delta q(r) \approx \frac{m^2r^2}{g^2} K_2\left(\frac{mr}{g} \right)}
\end{equation}
after substituting $M=\frac{m}{g}$ back in.
\end{document}

%% file: preamble.tex
\usepackage{amsfonts}
\usepackage{amscd}
\usepackage{amssymb}
\usepackage{amsmath,bbm}
\usepackage{graphicx}
\usepackage{subcaption}

\usepackage{epsfig}
\usepackage{latexsym}
\usepackage{mathtools}
\usepackage{hyperref} 
\usepackage[vcentermath]{youngtab}
\usepackage{mathrsfs}
\usepackage{url}
\usepackage{comment}
\usepackage{tikz}
\usepackage{tikz-cd}
\usetikzlibrary{arrows,arrows.meta}

\usepackage{pgfplots}

\usepackage{amsmath}
\usepackage{mathtools}
\usepackage{slashed}
\usepackage{braket}

\usepackage{tikz-feynman}

\hypersetup{colorlinks=true, linkcolor=blue,citecolor=blue,filecolor=black,urlcolor=black,}
    
\tikzcdset{arrow style=tikz, diagrams={>=stealth}}

\def \be  {\begin{equation}}
\def \ee  {\end{equation}}
\def \bea {\begin{equation}\begin{aligned}}
\def \eea {\end{aligned}\end{equation}}
\def \ba  {\begin{eqnarray}}
\def \ea  {\end{eqnarray}}
\def \bb  {\begin{thebibliography}}
\def \eb  {\end{thebibliography}}
\def \lab #1 {\label{#1}}

\newcommand\cA{\mathcal{A}}

\newcommand\cD{\mathcal{D}}

\newcommand\cL{\mathcal{L}}

\newcommand\cO{\mathcal{O}}

\newcommand\cU{\mathcal{U}}

\newcommand\cW{\mathcal{W}}

\newcommand\bQ{\mathbb{Q}}

\newcommand\bZ{\mathbb{Z}}

\newcommand\vect{\mathsf{Vec}}

\newcommand\rep{\mathsf{Rep}}

\definecolor{cardinal}{rgb}{0.6,0,0}
\definecolor{darkgreen}{rgb}{0,0.5,0}
\definecolor{golden}{rgb}{0.92, 0.7, 0}
\definecolor{midnight}{rgb}{0, 0, 0.5}
\definecolor{darkblue}{rgb}{0.2, 0, 0.8}